# Joint Proportional Fairness Scheduling Using Iterative Search for mmWave Concurrent Transmission

Ahmed M. Nor
*Telecommunications Department.*
*University Politehnica of*
*Bucharest*Bucharest, Romania
*Electrical Engineering Department*
*Aswan University*
Aswan, Egypt
ahmed.nor@upb.ro

***Abstract*—Millimeter wave (mmWave) will play a significant role as a 5G candidate in facing the growing demand of enormous data rate in the near future. The conventional mmWave standard, IEEE 802.11ad, considers establishing only one mmWave link in wireless local area network (WLAN) to provide multi Gbps data rate. But, mmWave has a tenuous channel which hinders it from providing such rate. Hence, it's necessary to establish multiple mmWave links simultaneously by deploying a multiple number of mmWave access points (APs) in 5G networks. Unfortunately, applying conventional standard without any modifications for mmWave concurrent transmission impedes mmWave APs from selecting optimum mmWave concurrent links. Because IEEE 802.11ad standard associates the user equipment (UEs) to mmWave APs using the link that has the maximum received power without considering mutual interference between simultaneous links. In this paper, a joint proportional fairness scheduling (JPFS) optimization problem for establishing optimum mmWave concurrent transmission links is formulated. However, it is highly complicated to find a solution to this non-polynomial (NP) time problem using exhaustive search. Hence, an iterative search (IS) scheme is proposed to obtain a sub-optimal solution to it with highly relaxing its complexity. Numerical simulation proves the effectiveness of using the proposed IS scheme to improve the system performance and reduce the system complexity.**

## I. Introduction

The intensive need of enormous data rate and low latency connection will be a big issue in the upcoming years due to the rapidly growth in using high speed mobile networks, cloud computing, and IoT applications [1], [2]. To overcome on this issue, 5G networks are widely deployed nowadays specially in indoor areas. Millimeter wave (mmWave) is one of the candidates to enable 5G network hence it becomes a hot research topic particularly 60 GHz band [3], [4]. Although the availability of wide unlicensed spectrum, mmWave band faces multiple technical issues due to the channel nature in such as highly propagation and penetration losses [3]. Directional communication based on antenna beamforming gains is adopted by the mmWave standard to enable high speed mmWave transmission links [5]. However, mmWave links still have a high probability to be blocked due to human shadowing [6], [7]. Moreover, indoor areas, for instance airports and stadiums are crowded by multiple user equipment (UEs) where each UE needs high data rate link. That means 5G networks using only one mmWave access point (AP) cannot achieve the required multi Gbps data rate. Hence, network should contain multiple mmWave APs to enable mmWave concurrent transmission by scheduling multiple mmWave links at the same time slot. However, these concurrent links suffer from highly mutual interference which incredibly reduces the system throughput. To mitigate mutual interference between links, network shall select the optimum links that can be scheduled at the same time slot [8].

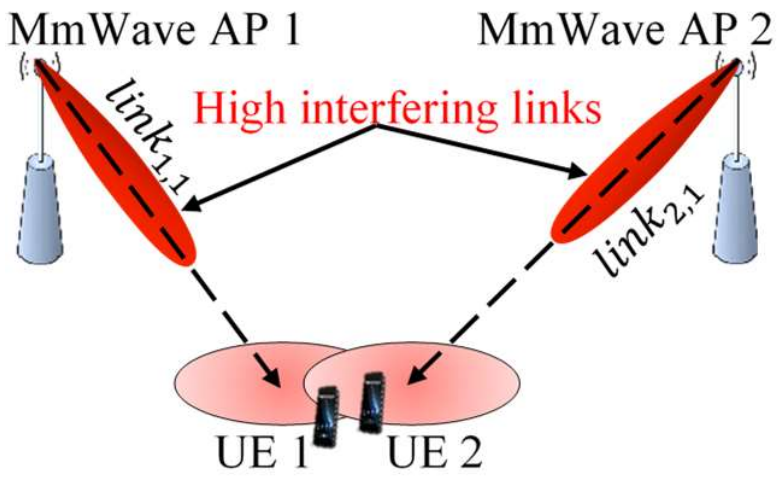

Fig.1. Non-optimal establishment of mmWave concurrent links based on IEEE 802.11ad standard.

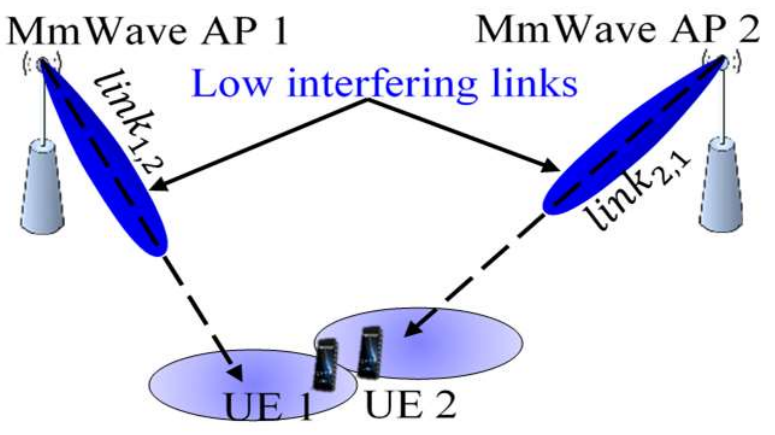

Fig. 2. Optimal establishment of mmWave concurrent links.

To cope with this need, IEEE 802.11 ad standard with scheduled medium access control (MAC) can be applied for establishing mmWave concurrent transmission [9]. But it cannot guarantee selecting optimal mmWave links. Because it selects each concurrent link separately based on the maximum received power criteria without considering the interference generated from other connected mmWave links. For example, as shown in Fig.1, IEEE 802.11 ad standard based WLAN selects $link_{1,1}$ and $link_{2,1}$ to establish mmWave concurrent transmission for UE1 and UE2, respectively. But these two links are high interfering links because there is high mutual interference between them. On the other side, $link_{1,2}$ and $link_{2,1}$ are low interfering links as shown in Fig.2, hence using them guarantees higher total system data rate. Thus, if WLAN takes mutual interference into consideration, these low interfering links can be selected as mmWave concurrent link. Moreover, using scheduled MAC with round robin (RR) scheduling mechanism in each mmWave AP separately to allocate time resources to UEs, cannot guarantee efficient resource allocation hence wasting the opportunity to benefit from all available capacity. Besides, UEs with highly correlated channel might be selected to connect with their associated mmWave APs simultaneously which increase the probability of mutual interference occurrence between concurrent links hence reducing the total system data rate. So, selecting the optimal mmWave concurrent links is a must.

In this paper, a joint proportional fairness scheduling (JPFS) optimization problem for mmWave concurrent transmission is formulated with considering the mutual interference between simultaneous links. This problem aims to select optimal concurrent links that maximizing the total system data rate by jointly performing user association and time resource allocation. This JPFS optimization problem is a non-polynomial (NP) time problem particularly when network consists of multiple number of mmWave APs and UEs. Hence, finding its optimum solution using exhaustive search (ES), i.e., searching on all available mmWave APs-UEs patterns in exhaustive manner, is complicated. Thus, we propose an iterative search based JPFS (IS-JPFS) algorithm to find a sub-optimal solution to this problem. Where concurrent links are selected link by link with considering the already connected links interferences in an iterative manner until achieving a convergence point instead of searching on all available APs-UEs patterns to select the optimum one. Hence, highly relaxing the ES-JPFS complexity, reducing its total setup time, and putting a polynomial time solution to it. Besides, obtaining nearly the same ES-JPFS performance.

This paper is organized as follows: Sec. II shows the proposed system model. Then, JPFS optimization problem for mmWave concurrent transmission is formulated in Sec. III. The proposed iterative search based JPFS scheme is presented in Sec. IV, then the numerical analysis is discussed in Sec. V. Finally, Sec. VI presents the conclusion of the paper.

## II. Proposed System Model

The proposed network architecture is shown in Fig.3. This network consists of $M$ mmWave APs, $K$ users and an access point controller (APC). Highly traffic indoor areas such as airports and stadiums are examples where this network can be installed. High speed back-hull links, e.g., fibre optic, are used to connect network component, i.e., mmWave APs and APC. In this network, APC utilized to jointly associate and scheduled UEs to mmWave APs by establishing the optimal (sub-optimal) mmWave concurrent transmission links that can be scheduled simultaneously. Besides, APC considered as a gateway between WLAN and the Internet.

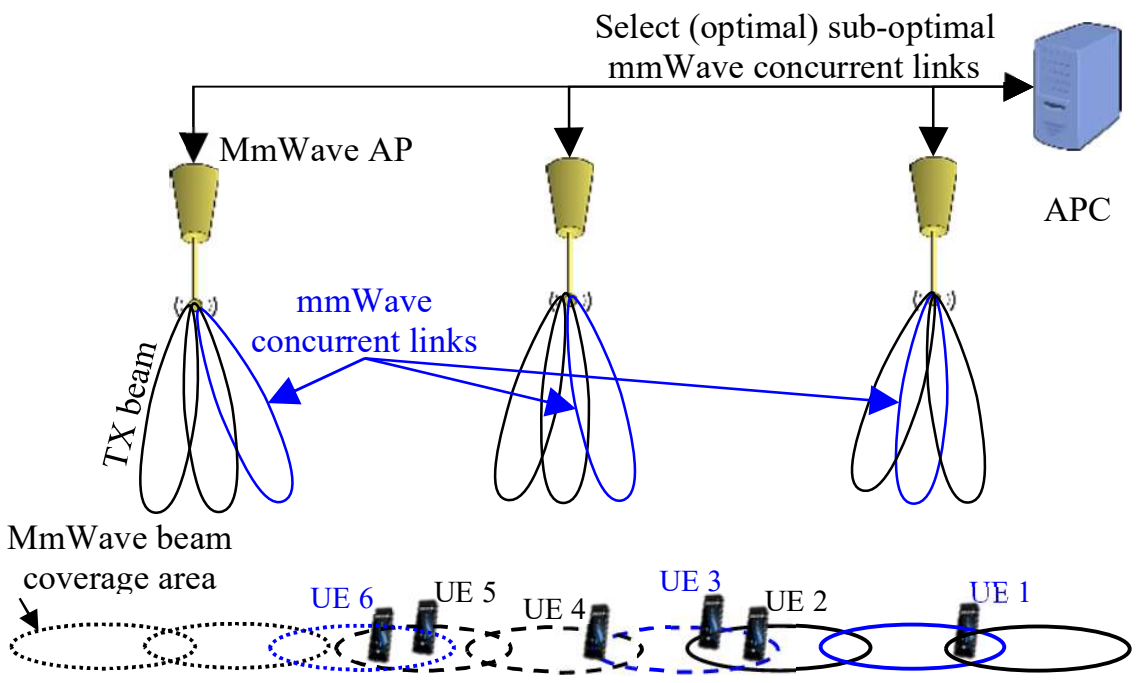

Fig. 3. Proposed WLAN network.

For simplicity, we considered only beamforming in mmWave APs while considering quasi-omni antenna model for UEs. The instantaneous achievable data rate $r_{mk}(n)$ at user $k$ from mmWave AP $m$ at time slot $n$, presented as:

$$r_{mk}(n) = \eta_{BW} BW \log_2\left(1 + \eta_{SNR} SNR_{mk}^{b_m}(n)\right), \quad (1)$$

where $\eta_{BW}$ and $\eta_{SNR}$ are the bandwidth efficiency and the SNR efficiency, respectively. $BW$ indicates to the bandwidth, while $SNR_{mk}^{b_m}(n)$ indicates to the SNR for user $k$ in case of connection with AP $m$ by beam ID $b_m$ and can be written as:

$$SNR_{mk}^{b_m}(n) = \frac{Pr_{mk}^{b_m}(n)}{N_o}, \quad (2)$$

where $N_o$ donates to the noise power. $Pr_{mk}^{b_m}$ is the received power of user $k$ which is far from mmWave AP $m$ by distance $d$ and connected to it using transmitted beam ID $b_m$. Hence, this transmitted power can be calculated by using mmWave pathloss model and written as [5]:

$$Pr(d)[dB] = G(\varphi,\theta) + PL(d) + P_t, \quad (3)$$

$$PL(d)[dB] = PL_o(d_o) + 10 n_{exp} \log_{10}\left(\frac{d}{d_o}\right) + \chi_\sigma, \quad (4)$$

where $P_t$ indicates the transmitted (TX) power for mmWave in dBm, $PL(d)$ refers to the mmWave average path loss in dB and $PL_o(d_o)$ is the path loss at a reference distance $d_o$. $n_{exp}$ and $\chi_\sigma$ indicate to the mmWave path loss exponent and shadowing term, respectively. Where $\chi_\sigma \backsim \mathcal{N}(0, \sigma_{\chi_\sigma})$ as $\sigma_{\chi_\sigma}$ is the standard deviation of $\chi_\sigma$ . $G(\varphi,\theta)$ indicates the 3D beamforming model for mmWave gain of TX beam in dB. Hence, $G(\varphi,\theta)$ can be written as [5]:

$$G(\varphi,\theta) = G_o[dB] - \min\left[-\left(G^H(\varphi), +G^V(\theta)\right), \Upsilon\right], \quad (5)$$

$$\Upsilon = 12 + G_o[dB], \quad (6)$$

$$G_o[dB] = 20\log_{10}\left(\frac{1.6162}{\sin\left(\frac{\theta_{-3dB}}{2}\right)}\right), \quad (7)$$

where $\varphi$ and $\theta$ are the azimuth and the elevation angles, defining the angle of departures (AoDs) of mmWave channel, while $G^H(\varphi)$, $G^V(\theta)$ are the gains of TX beam in horizontal and vertical directions, that can be expressed as [5]:

$$G^H(\varphi) = -\min\left(12\left(\frac{\varphi - \varphi_{TX}}{\varphi_{-3dB}}\right)^2, \Upsilon\right), \quad (8)$$

$$G^V(\theta) = -\min\left(12\left(\frac{\theta - \theta_{TX}}{\theta_{-3dB}}\right)^2, \Upsilon\right), \quad (9)$$

where $\varphi_{TX}$ and $\theta_{TX}$ are the centres angles of mmWave AP TX beam, while $\varphi_{-3dB}$ and $\theta_{-3dB}$ are the beamwidths of mmWave beam in the vertical and the horizontal directions.

## III. Proplem Formulation

In this section the JPFS optimization problem is formulated. The optimization problem aims to select the optimum APs-UEs configuration pattern that can be scheduled concurrently to obtain the target cost function that maximizes the total system rate over users' average data rate $\overline{R_k}$ and considering the fairness between users. Hence, this problem can be formulated as [10]:

$$\max \sum_{k \in K} \log(\overline{R_k}) \tag{10}$$
$$\text{Subject to } \overline{R_k} \in R^+,$$

where $K$ is the UEs total number that located in the network, and $k$ is the index of a certain UE. While $R^+$ indicates the vectors of all users' achievable data rates.

The mmWave has a spatial diversity characteristic due to its directional transmission and its standard use time division multiple access (TDMA) as a multiple access technique. The JPFS optimization problem shall be solved by APC at each time slot $n$. Thus, it converted into a per time slot AP-UE assignment problem where $w(n) = (w_{mk}(n): m \in M, k \in K)$ is the indicator matrix for the AP-UE assignment with size $M \times K$ at time slot $n$. $w_{mk}(n) = 1$, means user $k$ associate with AP $m$ at time slot $n$, and $w_{mk}(n) = 0$ elsewhere.

Assuming that the number of UEs $K$ is larger than the number of APs exist in WLAN where indoor environments are crowded by users. Hence, (10) can be reformulated as a per time slot AP-UE assignment problem as:

$$\textbf{Assignment: } \max_{w(n)} \sum_{k \in K} \log\left(\overline{R_k}(n)\right) \tag{11}$$
$$\text{Subject to } \sum_{k \in K} w_{mk}(n) = 1, \forall m \in M.$$

where $\sum_{k \in K} w_{mk}(n) = 1, \forall m \in M$ means only single UE can be associated to mmWave AP $m$ at time slot $n$. While, $\overline{R_k}(n)$ indicates the average achievable data rate that given to user $k$ up to time slot $n$ which can be written as [9]:

$$\overline{R_k}(n) = \left(1 - \frac{1}{n}\right)\overline{R_k}(n-1) + \frac{1}{n} R_k(n), \tag{12}$$

where $R_k(n) = \sum_{m \in M} w_{mk}(n)\, r_{mk}(n)$ is the instantaneous rate that obtained by user $k$ at time slot $n$, and $r_{mk}(n)$ is the user $k$ instantaneous achievable rate when UE, $k$ associated with AP $m$ at time slot $n$.

Following Taylor series expansion:

$$\log\left(\overline{R_k}(n)\right) \approx \log\left(\overline{R_k}(n-1)\right) + \frac{1}{\overline{R_k}(n-1)}\left(\overline{R_k}(n) - \overline{R_k}(n-1)\right) \tag{13}$$

Thus using (12), (13) can be expressed as:

$$\log\left(\overline{R_k}(n)\right) \approx \log\left(\overline{R_k}(n-1)\right) + \frac{1}{n}\left(\frac{R_k(n)}{\overline{R_k}(n-1)} - 1\right), \tag{14}$$

Hence, maximizing $\log\left(\overline{R_k}(n)\right)$ is equivalent to maximize the preferred metric $\frac{R_k(n)}{\overline{R_k}(n-1)}$. Hence, the solution of the optimization problem can be expressed as:

$$w_{mk}(n) = \begin{cases} 1, & \text{if } \forall k \in K_M = arg \max_{\forall K_M \subset K} \left\{ \sum_{k \in K_M} \left(\frac{R_k(n)}{\overline{R_k}(n-1)}\right) \right\}, \\ 0, & elsewhere. \end{cases} \tag{15}$$

where $\forall K_M \subset K$ means all users configurations of length $M$ of all network users $K$. This solution indicates that the APC selects the user pattern $K_M$ that obtain the maximum of $\sum_{k \in K_M}\left(\frac{R_k(n)}{\overline{R_k}(n-1)}\right)$ to be scheduled in time slot $n$, hence for $\forall k \in K_M$ the AP-UE assignment indicator $w_{mk}(n)$ will be equal to 1. Assuming the network consist of single mmWave AP, i.e., $M = 1$, (15) will become:

$$w_{mk}(n) = \begin{cases} 1, & if\ k = arg \max_{\forall k \subset K} \left(\frac{R_k(n)}{\overline{R_k}(n-1)}\right), \\ 0, & elsewhere. \end{cases} \tag{16}$$

## IV. Proposed MmWave Concurrent Transmission

The optimization problem that formulated in section III, is a NP-hard problem. To find the optimum solution to this optimization problem as given in (15), APC needs to exhaustively search on all available APs-UEs configuration pattern using all available mmWave AP beams to find the optimum pattern maximizes the total system data rate. This solution needs a complexity equals to $\frac{K!}{(K-M)!}B^M$, where B indicates the number of mmWave AP transmitter (TX) beams. This solution complexity increases non-linearly with the number of deployed mmWave APs on the network and the number of UEs desire to connect with a high-speed mmWave concurrent links. Therefore, this exhaustive search based JPFS (ES-JPFS) scheme needs a high-power consumption and a long setup time specially in indoor areas which is the use case under study in this chapter. To overcome on the complexity issue, an iterative search scheme is proposed to find the sub-optimum solution of the aforementioned JPFS optimization problem. This iterative search based JPFS (IS-JPFS) scheme highly relaxes the optimization problem complexity and provide a polynomial time solution for it.

In the proposed IS-JPFS, instead of selecting the optimum APs-UEs pattern directly by searching on all available patterns, the mmWave concurrent links are selected link by link. Where APC selects each AP-UE link considering the interference produced by the links that have been already selected before. This producer is performed iteratively till obtaining a convergence point. In our proposal, the convergence point is obtained when the achievable rate obtained by selecting APs-UEs pattern in a certain iteration $R_{iter}$, is smaller than or equal to the achievable rate achieved by the APs-UEs pattern that selected in the previous iteration $R_{iter-1}$ by a factor of $\Delta_{th}$, i.e., $R_{iter} \leq \boldsymbol{\Delta_{th}} R_{ite}$ . The threshold $\boldsymbol{\Delta_{th}}$ is a design parameter determined based on the required performance and considering the acceptable complexity from the IS-JPFS scheme. In this chapter, the threshold $\boldsymbol{\Delta_{th}}$ selected to be equal to 0.1 but this parameter will be put under study in our future work.

The proposed IS-JPFS scheme select the sub-optimum APs-UEs pattern not the optimal one as in the ES-JPFS and maintains nearly the same performance that can be obtained by using the NP-hard solution but highly relaxes the system complexity. The proposed IS-JPFS scheme focuses only on the problem of selecting the sub-optimal APs-UEs pattern without considering the issue of selecting an efficient beam pattern. Because there are efficient mmWave beamforming training schems that proposed in literature as in [2], [11].

The proposed scheme aims to select the sub-optimal APs-UEs pattern at every time slot that maximizes the total system data rate and obtaining (15). Selecting the sub-optimal APs-UEs pattern means assigning UE $k$ index and beam identification $b_m$ to the user sets $K_M$ and the beam sets $B_M$ of all $M$ links at every time slot $n$. Thus, the proposed IS-JPFS explained as follows: Firstly, $K_M$ and $B_M$ are initialized as $K_M = \vec{0}$ and $B_M = \vec{0}$ with length $M$. In the first iteration, the first link is selected by performing BT for single link, i.e., considering there is no interference will be generated from concurrent links. Thus, $K_M[1]$ and $B_M[1]$ are assigned by selected UE and beam that obtained (15).

Regarding to the following $m$th links where $1 < m \leq M$, the APC performs BT for each link $m$ with considering the interference produced from the links that have been selected before. Hence, $K_M[m]$ and $B_M[m]$ can be updated with the UE and beam that obtained by (15). The first iteration is ended when APC selects all $M$ concurrent links. In the next iterations, the selection of each link $m$ considers the interference generated from the other $M-1$ links since the user sets $K_M$ and the beam sets $B_M$ have been already determined before. Hence, in each iterative round, the data rate obtained from the modified selected links is larger than or equal to that obtained from the previously selected links. So, the process of concurrent links selection is repeated until obtaining a convergence point, i.e., $R_{iter} \leq \Delta_{th} R_{iter-1}$. Algorithm 1 summarizes the proposed IS-JPFS.

## V. Simulation Results

In this numerical simulation, the performance of the proposed IS-JPFS scheme for mmWave concurrent transmission is compared with ES-JPFS and the conventional scheme, i.e., IEEE 802.11ad based WLAN, under different scenarios. Table. I. shows the main simulation parameters used in the numerical analysis. Where a medium conference room, which is shown in Fig. 4., is selected as a target area. Also, the human shadowing blocking probability assumed to be 0.5 for the mmWave line of sight path [6], [7]. The number of mmWave APs is 6 which is considered suitable number of APs in such indoor area to provide required data rates to users. A mmWave beamwidth of 30° is used through this study as an example. In the conventional scheme, UEs are associated to mmWave APs based on the maximum received power while each mmWave AP scheduled its associated UEs separately based on scheduled MAC as defined by IEEE 802.11ad standard. On the other side, for the proposed IS-JPFS and ES-JPFS, mmWave UEs are jointly associated and scheduled to mmWave APs in each time slot.

**Algorithm 1: IS-JPFS for mmWave concurrent transmission**

**Input:** $\Delta_{th}$
**Start**

1. **Initialize** the user sets $K_M = \vec{0}$, the beam sets $B_M = \vec{0}$, and $R_{iter} = 0$.
2. **Repeat**
3. **Assign** $R_{pre} = R_{iter}$.
4. **For** AP $m = 1$ to $M$ ($m$ is AP number index) **do**
5. **set** $K_M$ and $B_M$ according to
6. $arg \max_{\forall K_M \subset K} \left\{ \sum_{k \in K_M} \left( \frac{R_k(n)}{\overline{R_k}(n-1)} \right) \right\}$,
7. **end**
8. **Until** $R_{iter} \leq \Delta_{th} R_{pre}$

**Stop**
**Output:** $\boldsymbol{K_M}$, and $\boldsymbol{B_M}$

Table. I: Simulation Parameters.

| Parameter | Value |
|---|---|
| Room dimensions | 20m × 10m × 4m |
| UEs height from floor | 1m |
| Number of mmWave APs / UEs | 6 / 40 |
| TX power of mmWave AP | 10 dBm |
| MmWave Beamwidth | 30° |
| MmWave Bandwidth | 2.16 GHz |
| Bandwidth efficiency $\eta_s$ | 0.6 |
| SNR efficiency $\eta$ | 1 |
| Blocking probability | 0.5 |

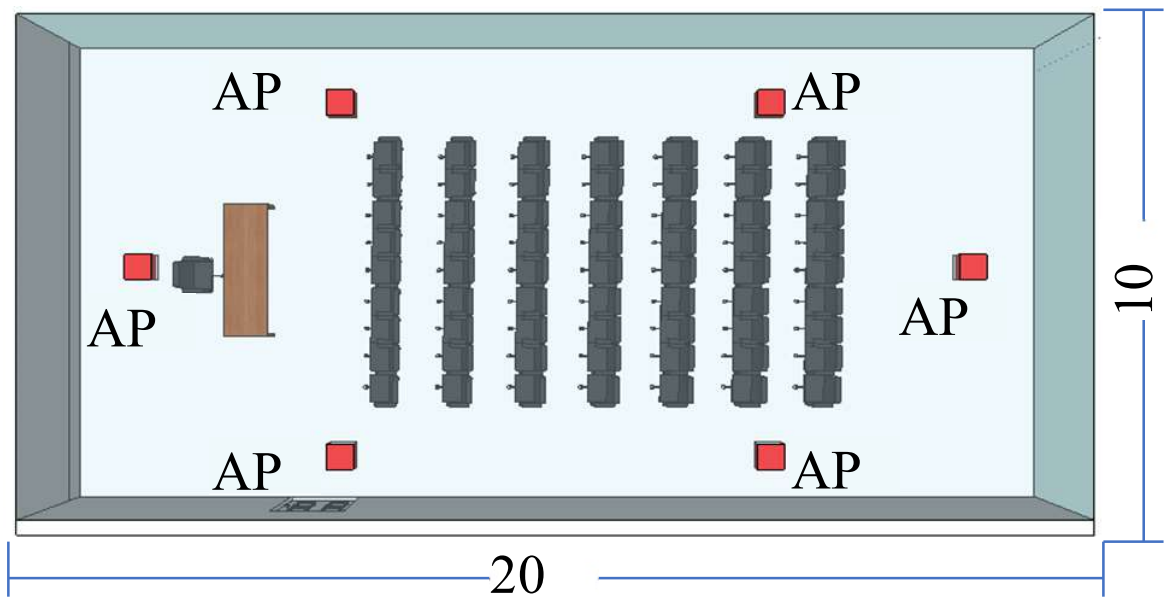

Fig. 4. Study area.

### A. Performance Metrics

In these numerical simulations, we focus on measuring the total system data rate in Gbps. The spatial reuse factor $\rho$ is used as an indicator for the obtained rate to the average rate in case of full isolated links and can be written as [12]:

$$\rho = \frac{\text{Sum rate of mmWave concurrent links}}{\text{Average rate of isolated links}} \tag{17}$$

The large value of $\rho$ means low mutual interference between links and efficient spatial reuse [10], e.g., if 6 mmWave APs exists in WLAN with no mutual interference between their links, $\rho$ will equal to 6.

Also, users' fairness index (FI) is evaluated to measure the variance between data rates provided to different users and can be expressed as:

$$FI = \left( \sum_{k=1}^{K} r_k \right)^2 \Big/ \left( K \sum_{k=1}^{K} {r_k}^2 \right) \tag{18}$$

where $r_k$ is the data rate offered to user $k$ a long all time slots $N$ from all mmWave APs and can be expressed as:

$$r_k = \sum_{m=1}^{M} \sum_{n=1}^{N} r_{mk}(n), \tag{19}$$

Moreover, the total system computational complexity in number of beams switchings is presented for the three schemes, where the complexity of the conventional IEEE 802.11ad based WLAN can be expressed as:

$$C_{conv.} = BM, \tag{20}$$

where APC synchronize between mmWave APs, so AP begins its BT process on all its beams $B$ when the previous AP ends it BT process. While the ES-JPFS computational complexity can be written as:

$$C_{ES-JPFS} = \frac{K!}{(K-M)!} B^M, \quad (21)$$

Where APC searches on all available APs-UEs configuration patterns to find optimum one. Regarding to the proposed IS-JPFS scheme, the complexity can be defined as:

$$C_{IS-JPFS} = \alpha\, B\, M, \quad (22)$$

where $\alpha$ indicates to the average number of iterations performed by the proposed IS-JPFS scheme to obtain the convergence point which has been defined in the Section IV.

*B. Results*

Fig. 5, 6 and 7 show the performance of the proposed IS-JPFS, ES-JPFS, and IEEE 802.11ad based WLAN, i.e., the conventional scheme, when different number of mmWave APs are used for mmWave concurrent transmission at a certain user density which is 1 user/m$^2$. While, Fig. 6 presents the comparison between the three schemes in terms of the total system complexity in number of beams switchings.

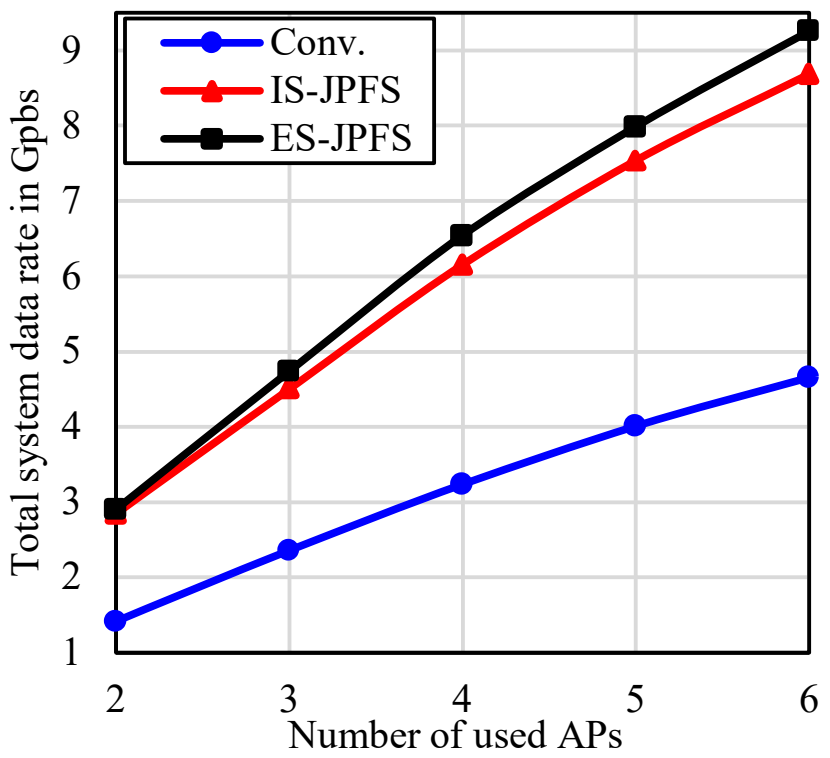

Fig. 5. Total system data rate using different number of mmWave APs.

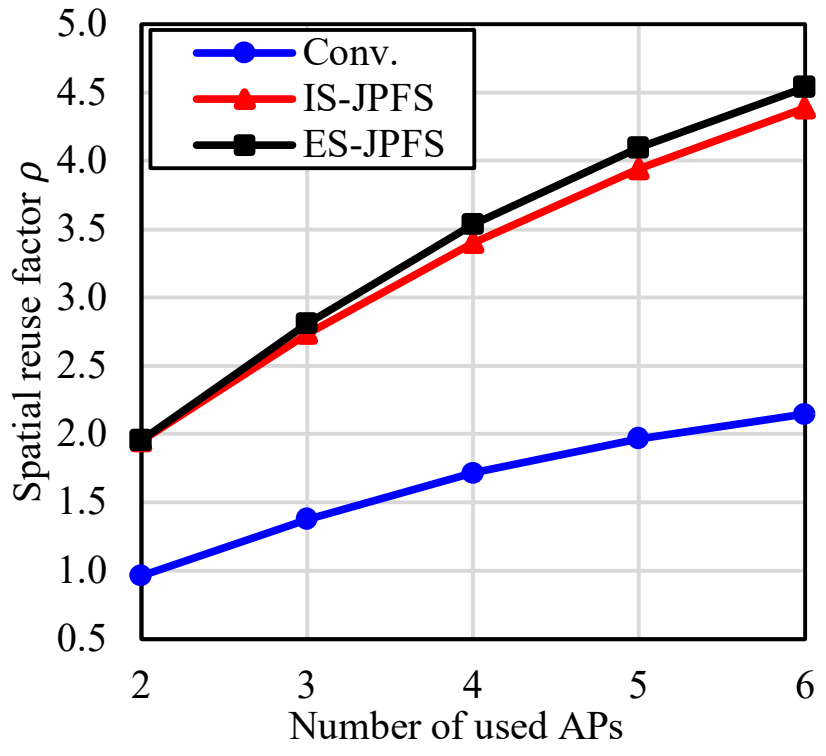

Fig. 6. Spatial reuse factor using different number of mmWave APs.

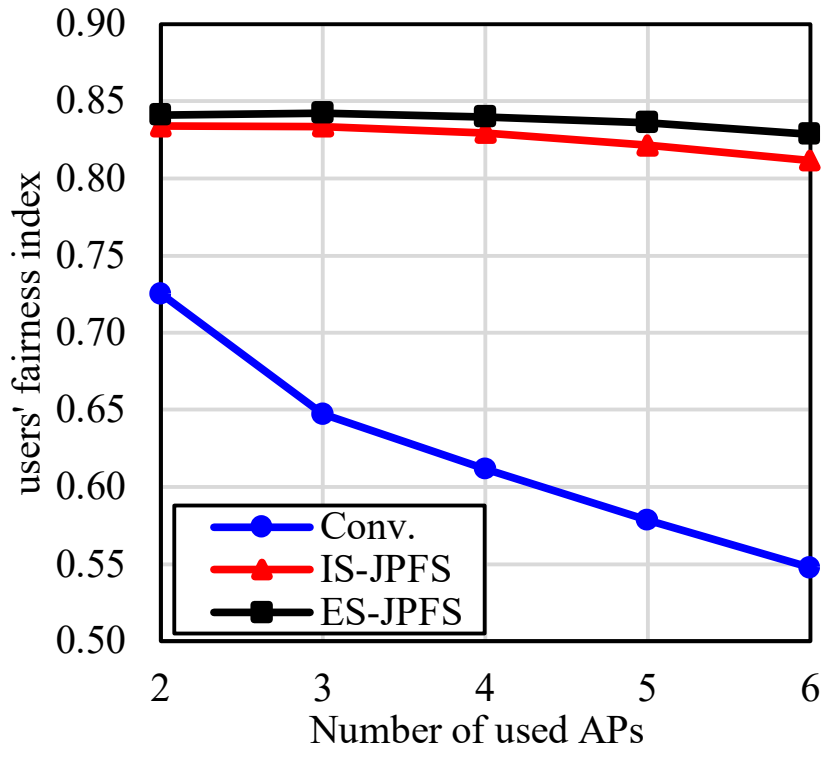

Fig. 7. User's fairness index using different number of mmWave APs.

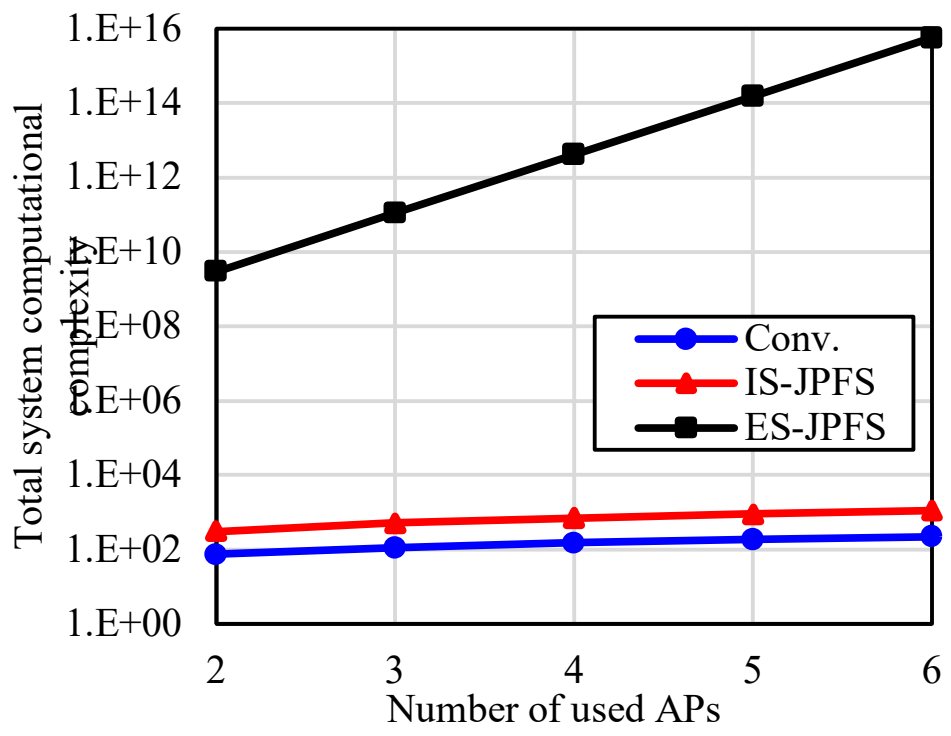

Fig. 8. System complexity when number of mmWave APs are used.

As shown in Fig. 5, 6 and 7, the performance achievable by ES-JPFS outperform the conventional scheme in terms of the total system data rate, spatial reuse factor $\rho$ and user's fairness index as a result of optimization problem where interference between concurrent links is mitigated. For instance, in case of 6 mmWave APs are used, the proposed ES-JPFS outperform the conventional scheme by 2.1-time increase in total system date rate, a 2-time increase in spatial reuse factor and nearly a 1.5-time increase in the user's fairness index. However, the ES-JPFS significantly increases the total system complexity comparable to the conventional scheme as shown in Fig. 6. Hence, the IS-JPFS scheme is proposed to overcome this complexity with acceptable performance. As presented in Fig. 8, the proposed IS-JPFS algorithm highly reduces the total system complexity comparable to ES-JPFS particularly for large numbers of mmWave concurrent links. Besides, it achieves nearly the same performance of it. For example, in case of 6 mmWave links are used, the reduction of total system complexity of the proposed IS-JPFS over using the ES-JPFS is more than 99%. In addition, the IS-JPFS scheme needs only 5-times the complexity of the conventional scheme.

Fig. 9, 10 and 11 show the comparison in performance between the three schemes at different user densities in terms of total system data rate, spatial reuse factor $\rho$, and users' fairness index, respectively. Where, we select only APs 2, 3, 4, and 5 for mmWave concurrent transmission because these APs are located close to each other hence there will be a high overlapping between their coverage areas that means high mutual interference between their concurrent links which clearly tackle the problem under study.

As shown in Fig. 9 and 10, when the user density rises, the total system rate and spatial reuse factor $\rho$ are declined due to the high increase in mutual interference. While, as a result of interference, the difference between users' data rates is decreased hence the variance between $r_k$ decreased which improves the fairness between users as presented in Fig. 11. For instance, using the conventional scheme with increasing the user density from 0.2 to 2 user/m$^2$ decreases total system data rate by 30%. Besides, a loss of 50% in spatial reuse factor form ideal value, where $\rho = 4$, is obtained at user density 0.2 user/m$^2$, and a loss of 61% is obtained at user density 2 user/m$^2$. On the other hand, using the proposed IS-JPFS increases total system date rate by 92% and 65% at user densities 0.2 user/m$^2$ and 2 user/m$^2$, respectively. Moreover, the proposed IS-JPFS nearly obtains the ideal value of spatial reuse factor at user density 0.2 $\mathrm{user/m^2}$ and only a loss of 30% from ideal value at user density 2 $\mathrm{user/m^2}$. Also, the IS-JPFS enhanced users' fairness index by 50% in comparison with IEEE 802.11ad based WLAN.

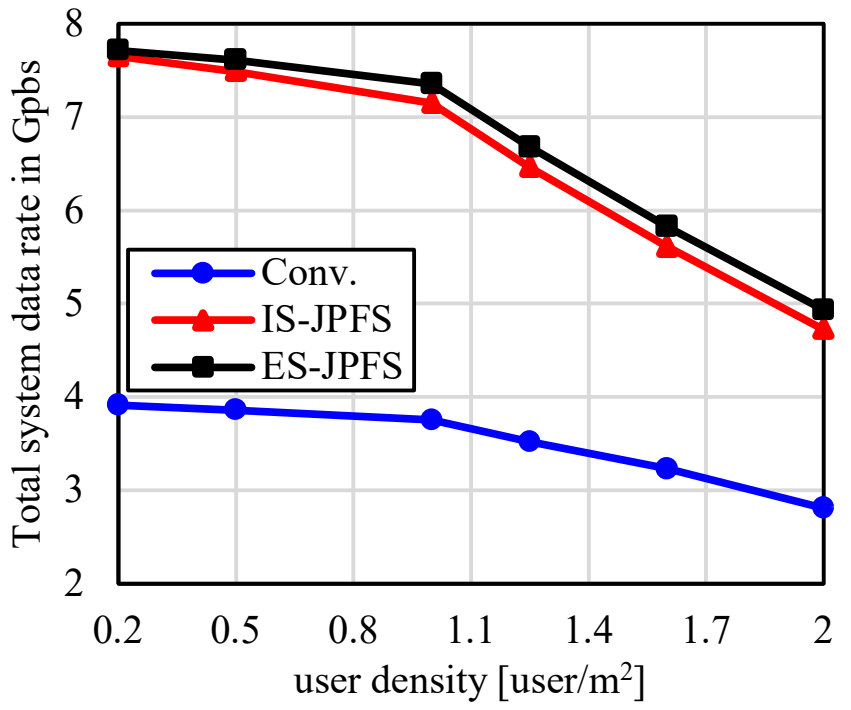

Fig. 9. Total system data rate in different user densities scenarios.

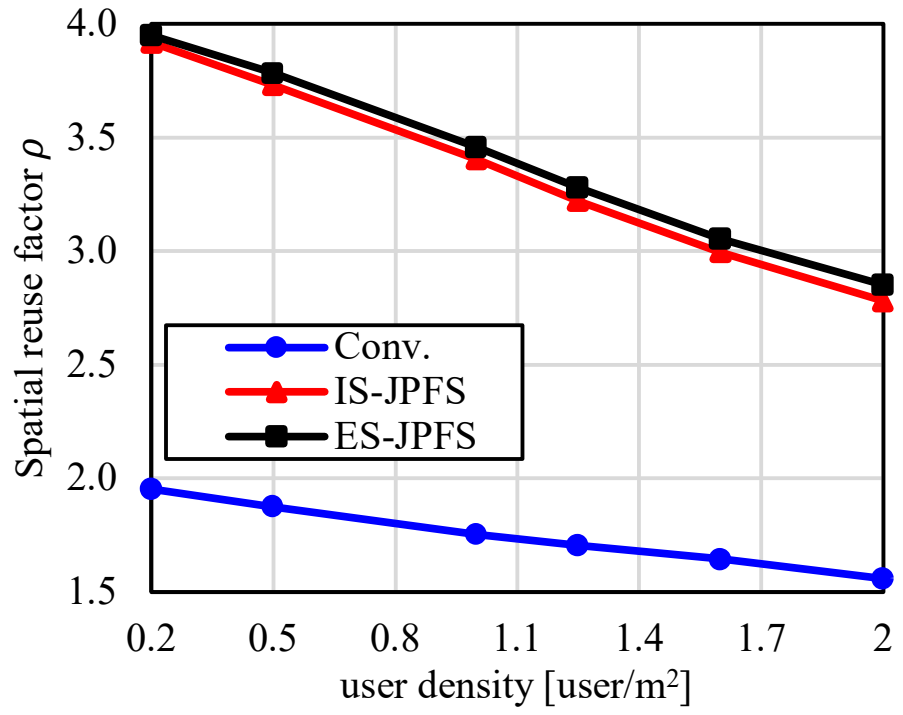

Fig. 10. Spatial reuse factor in different user densities scenarios.

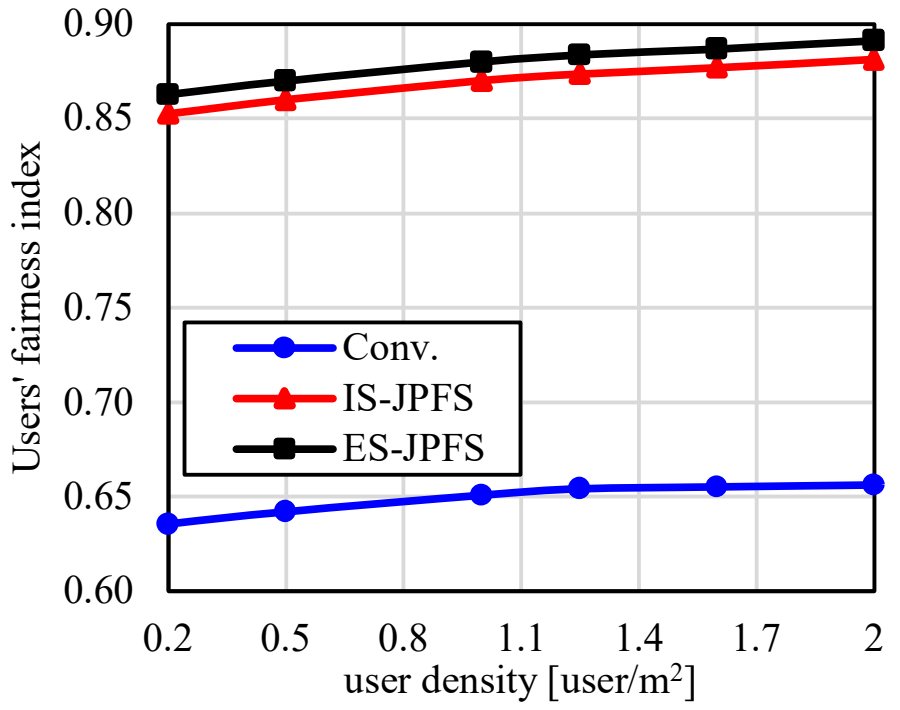

Fig. 11. Users' fairness index in different user densities scenarios.

## VI. Conclusion

In this paper, an optimization problem for mmWave concurrent transmission in scheduled MAC scenario is formulated. This problem aims to select the optimum mmWave links that can be scheduled at the same time. This JPFS optimization problem needs a non-polynomial time to find its solution using exhaustive search. Hence, an iterative search based joint proportional fairness scheduling scheme is proposed to highly relax the complexity of this problem. In which, APC selects a sub-optimal mmWave concurrent links in iterative manner, link by link, where each link considers the interference generated from the concurrent links that have been selected. According to numerical analysis, the proposed IS-JPFS scheme improves the system performance comparable to IEEE 802.11ad based WLAN in terms of total system data rate, spatial reuse factor $\rho$, and users' fairness index at the expense of increasing system complexity.

## Acknowledgment

This study has been conducted under the project 'MObility and Training fOR beyond 5G eco-systems (MOTOR5G)'. The project has received funding from the European Union's Horizon 2020 programme under the Marie Skłodowska Curie Actions (MSCA) Innovative Training Network (ITN) under grant agreement No. 861219.